\documentclass{ws-ijmpe}

\newcommand{\pt}{\ensuremath{p_{\mathrm{T}}}}
\def\ave#1{\langle {#1} \rangle}

\begin{document}

\markboth{Tapan K. Nayak}{Overview of event-by-event analysis}

\catchline{}{}{}{}{}
%%%%%%%%%%%%%%%%%%%%%%%%%%%%%%%%%%%%%%%%%%%%%%%%%%%%%%%%%%%%%%%%%%%%

\title{Overview of event-by-event analysis of high energy nuclear collisions}

\author{\footnotesize Tapan K. Nayak}

\address{CERN 
CH-1211, Geneva 23, Switzerland \\
and \\
Variable Energy Cyclotron Centre, Kolkata-700064, India \\
Tapan.Nayak@cern.ch}

\maketitle

\begin{history}
\received{(received date)}
\revised{(revised date)}
%\accepted{(Day Month Year)}
%\comby{(xxxxxxxxxx)}
\end{history}

\begin{abstract}
The event-by-event analysis of high energy nuclear collisions aims at
revealing the richness of the underlying event structures and 
provide unique measures of dynamical fluctuations associated with QGP phase transition.
The major challenge in these studies is to separate the dynamical fluctuations
from the many other sources which contribute to the measured values. We present the 
fluctuations in terms of event multiplicity, mean transverse momentum,
elliptic flow, source sizes, particle ratios and net charge distributions.
In addition, we discuss the effect of
long range correlations, disoriented chiral condensates and presence of jets.
A brief review of various probes used for fluctuation studies 
and available experimental results are presented.

\end{abstract}

\section{Introduction}

The event-by-event analysis of high energy nuclear collisions aims at searching for dynamical 
fluctuations associated with the phase transition of normal nuclear matter to the Quark-Gluon Plasma (QGP).
Fluctuations of thermodynamic quantities provide an unique framework
for studying the nature of QGP phase transition 
and provide direct insight into the properties 
of the system created in high energy heavy-ion collisions\cite{jeon-koch,heiselberg,nayak}.
Large fluctuations in energy density due to droplet 
formation\cite{vanhove} are expected if the phase transition is of first order. A
second order phase transition may lead to divergence in specific heat and increase
in fluctuations of energy density due to long range correlations in the system.
Furthermore, the prospect of locating the critical point of the QGP phase 
transition, where the fluctuations are predicted to be largely enhanced\cite{tricrit},
makes this study rather interesting and challenging.
The rapid development in the event-by-event study in recent years
is related to the availability of high beam energies and sophisticated
experiments with large acceptance detectors.
The regime of event-by-event study spans from understanding the bulk properties of matter
to high \pt~ particles including jets. 

The challenge of event-by-event studies is that, beyond the fluctuations linked to the 
details of the phase transition, there are a number of other fluctuations which appear. 
There are numerous well-established physical sources of event-by-event
fluctuations in high-energy nucleus--nucleus collisions, {\it viz.}, 
geometrical (impact parameter, number of participants, detector acceptance),
energy, momentum, temperature, charge conservations, anisotropic flow, 
Bose-Einstein correlations, resonance and string decays,
jets and minijets and effect of quantum statistics.
Many exotic phenomena
may also occur and significantly impact the observed fluctuations.
Among them are formation of Disoriented Chiral Condensates (DCC), colour collective phenomena 
and formation of colour ropes.

Fluctuations in physical quantities can shed light on the nature of the matter
created in relativistic heavy-ion collisions. Recently there has been a debate
over whether the bulk of the matter created at RHIC behaves like a perfect
fluid\cite{heinz}. Fluctuation in elliptic flow might provide
a sensitive probe towards answering this question.
Fluctuations of conserved quantities like net electric charge, baryon number and
strangeness are predicted to be significantly reduced in a QGP scenario
as they are generated in the early plasma stage of the system created in heavy-ion
collisions with quark and gluon degrees of freedom.
It has been suggested that the processes following QGP hadronization 
like hadronic rescattering
and resonance decays may almost completely wipe out fluctuations originally developed 
in the QGP phase. Thus the propagation of fluctuation from initial stages
of collision to the freeze-out has to be considered before making any 
conclusions about the fluctuations from QGP and non-QGP stages\cite{evolution}.

The information content of the amount of fluctuation is inherent in the variance of the width of the
distribution of a given observable, expressed in terms of 
\begin{equation}
     \omega_X = \frac{\sigma_X^2}{\langle X \rangle},
\label{eqn2}
\end{equation}
where $X$ is the variable under study,
$\sigma_X^2$ is the variance of the distribution 
and $\langle X \rangle$ denotes the mean value.
The task is to distinguish between statistical fluctuations and those which have
dynamical origin. Several methods have been put forward suggesting ways
to infer about the presence of dynamical fluctuations. In order to infer
about the presence of non-statistical fluctuations, one needs to compare the
experimental results with known models which incorporate all the known phenomena.
An alternate or may be complimentary procedure to probe the fluctuations in a model
independent manner would be to compare experimental distributions of real data to
those of the mixed events. In this manuscript, we discuss various probes of
fluctuations and recent experimental findings.

\section{Volume fluctuations and centrality selection}

The volume fluctuation\cite{jeon-koch} arises through the measurement of multiplicity, $N$,
\begin{equation}
        N =  \rho V,
\label{eqn0}
\end{equation}
where $\rho$ is the density and $V$ is the volume. The fluctuation in $N$ is expressed
as:
\begin{equation}
        \ave{\delta N^2} =  \ave{\delta \rho^2} \ave{V}^2 + \ave{\rho}^2 \ave{\delta V}^2.
\label{eqn00}
\end{equation}
Since the main interest is on the fluctuation of the density, $\ave{\rho}^2$, the second
term containing $\ave{\delta V}^2$ has to be estimated in order to make any conclusion.
One of the ways to control the volume fluctuation is by making proper centrality selection.

\begin{figure}
\begin{center}
\includegraphics[width=1.0\textwidth]{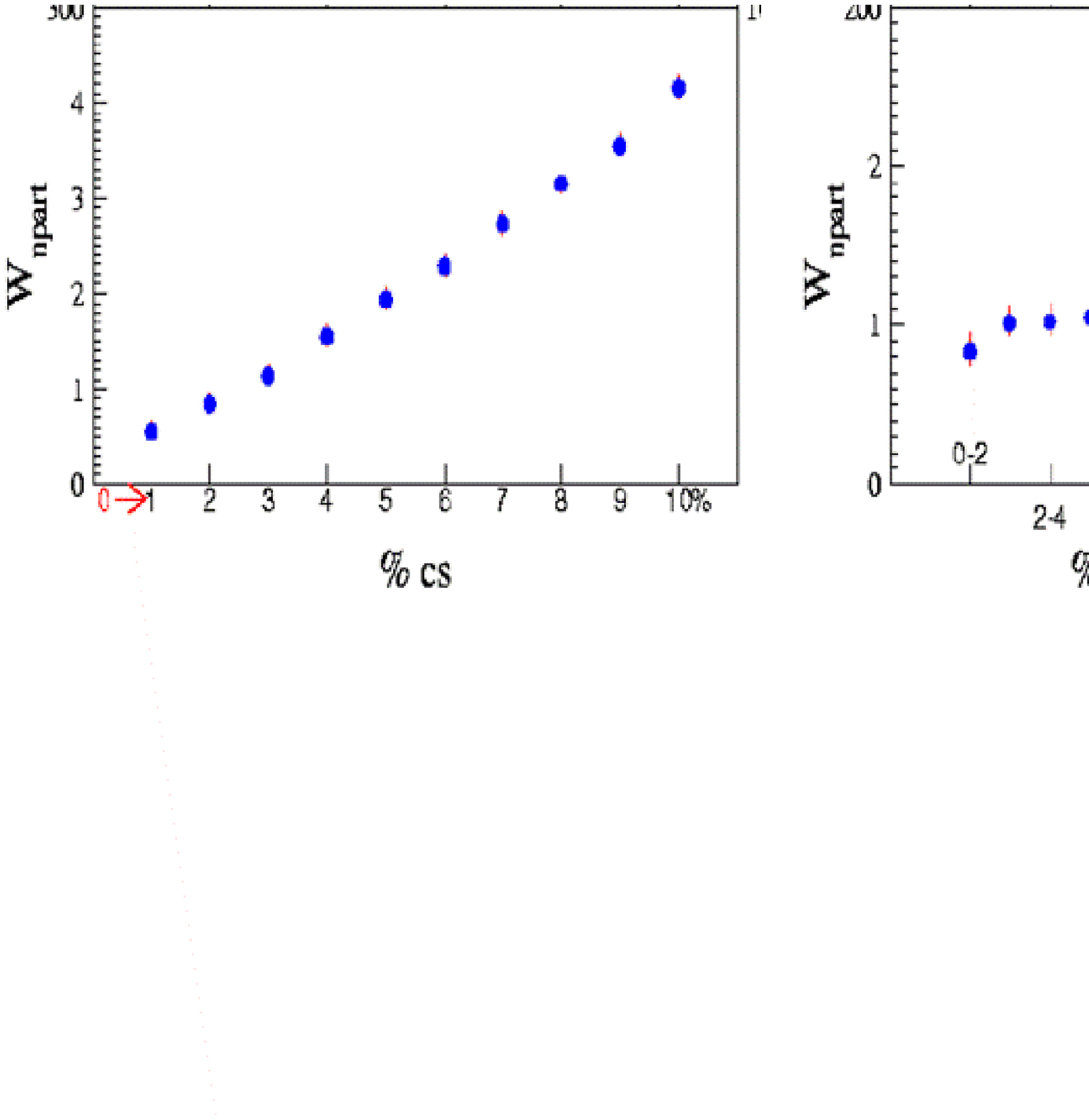}
\vspace*{-1.0cm}
\caption{
The fluctuation in the number of participants ($\omega_{npart}$) 
as a function of centrality,
expressed as a percentage of cross section for \mbox{Pb-Pb} 
collisions at $\sqrt{s_{\rm NN}}$=17.3~GeV. The increase of centrality bin
from very narrowly defined (0-1\%) to wider ones (0-10\%)
(shown in the left panel) causes an increase in the fluctuation.
The right
panel shows that for narrow centrality bins the fluctuations
remain minimal and close to unity.
}
\label{centrality}
\end{center}
\end{figure}

In case of heavy-ion collisions, centrality is characterized by the impact parameter, 
$b$, of the
collision, which also can be expressed in terms of the number of participating 
nucleons, $N_{\rm part}$. A given centrality class
has a set of values of $b$ or $N_{\rm part}$.
As there is no real control over the impact parameter
of the collision in heavy-ion experiments, geometric fluctuation is unavoidable in
the fluctuation of any extensive quantities\cite{jeon-koch,wa98_ebye}.
The importance of centrality selection for fluctuation studies can be understood
in terms of a participant model\cite{heiselberg,wa98_ebye,baym}.
Since it is not possible to measure either $b$ or $N_{\rm part}$ directly,
estimations of these quantities are based on calorimetric and
multiplicity measurements. 
For events in a given centrality class, $b$ or $N_{\rm part}$ values are extracted in a
model dependent way. 
The number of produced particles ($N$) in a collision depends on the centrality of the collision
expressed in terms of $N_{part}$ and the number of collisions 
suffered by each particle:
\begin{equation}
        N =  \sum_{i=1}^{N_{\mathrm part}}  n_i,
\label{eqn1}
\end{equation}
  where $n_i$ is the number of
  particles produced in the detector acceptance by the $i^{th}$ participant.
  The mean value of $n_i$ is the ratio of the 
  average multiplicity
  in the detector coverage to the average number of participants, i.e.,
  $\langle n \rangle = \langle N \rangle / \langle N_{\mathrm part} \rangle $. 
The fluctuation in particle multiplicity has a main contribution from 
the fluctuations in ($N_{\rm part}$). In order to infer any dynamical fluctuation
arising from various physics processes one has to make sure that the fluctuations in 
$N_{\rm part}$ are minimal. 

Fluctuations in $N_{\rm part}$ 
have been studied at the SPS by the WA98 experiment\cite{wa98_ebye} where the
centrality selections were made by using the mid-rapidity and the zero-degree
calorimeters. $N_{\rm part}$ values are calculated using the VENUS event 
generator\cite{venus} and the WA98 simulation framework.
Figure~\ref{centrality} shows fluctuations in $N_{\rm part}$ for various ranges of
centrality bins expressed in terms of percentage of cross section. 
Fluctuation seems to increase for broad centrality class as shown in the
left panel of the figure, whereas the fluctuations for narrow
centrality bins (such as  0--2\%, 2--4\%, 4--6\%, ....,50--52\%)
remain around unity for most of the centrality bins. 
This suggests narrow cross section slices in the centrality bins are 
preferable for fluctuation studies.

\section{Multiplicity fluctuations}

Depending on the nature of QGP phase transition, there will be large density
fluctuations leading to droplet formation and hot spots\cite{vanhove}. These will give rise to
large rapidity and multiplicity fluctuations of produced particles and have
distinct effects on the space time extent of the source.
Multiplicity of produced particles characterizes the evolving system in a 
heavy-ion collision and thus fluctuation in multiplicity may provide a distinct
signal of the QGP phase transition\cite{heiselberg,wa98_ebye}. 

\begin{figure} 
  \begin{center}
    \leavevmode
    \epsfig{file=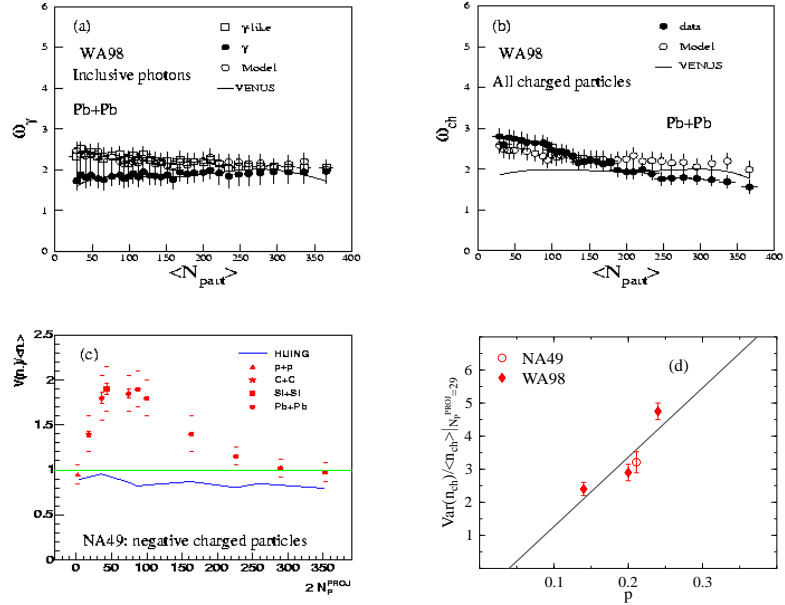, width=4.3in}
  \end{center}
\vspace*{-0.7cm}
  \caption{Multiplicity fluctuations at the SPS energies for 
(a) photons and (b) charged particles from the
WA98 experiment at the SPS, (c) charged particles from the NA49
experiment at the SPS and (d) comparison of the scaled variance of
charged particles for semi-central collisions as a function of 
acceptance for the WA98 and NA49 setup.
}
\label{mult_fluc}
\end{figure}
\begin{figure}[htp]
\begin{center}
\includegraphics[scale=0.64]{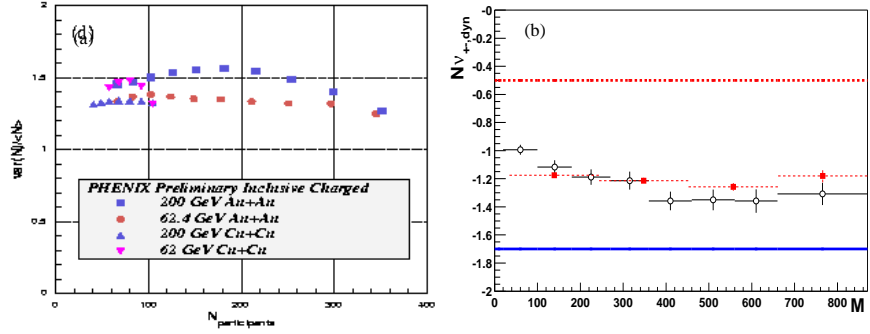}
\vspace*{-0.2cm}
\caption {\label{mult_fluc_rhic}
Multiplicity fluctuations of charged particles at the RHIC energies 
for PHENIX experiment (left panel) and STAR experiment (right panel).
}
\end{center}
\end{figure}
Since multiplicity distributions for narrow centrality bins can be described by Gaussian 
distributions, their fluctuations are expressed in terms of scaled variance, 
defined as, $\omega=var(N)/\ave{N}$, where $\ave{N}$ and $var(N)$ represent the variance and
mean of the multiplicity distribution, respectively.
Figure~\ref{mult_fluc} shows observed scaled variance for SPS and RHIC 
energies. The results from the WA98 experiment\cite{wa98_ebye}, those corresponding to
photons and charged particles, are compared to different model calculations.
Although the 
experimental data is consistent with the model calculations within the
quoted error bars, the increasing trend of fluctuation for charged particles
towards peripheral collisions is clearly visible. 
The scaled variance of charged particles as a function of centrality
as measured by NA49
experiment\cite{na49_mult} shows a non-monotonic behaviour, especially
at mid central regions. A good comparison has
been made between the results of the charged particles for WA98
and NA49 experiments by taking the acceptance and fraction ($p$) of registered
particles into account. As shown in Figure~\ref{mult_fluc}(d), the
results of both the experiments are in good agreement.
The PHENIX data\cite{phenix_mult,phenix_mult_fluc} for multiplicity fluctuations
are shown in the left panel of Figure~\ref{mult_fluc_rhic} for \mbox{Au--Au} and \mbox{Cu--Cu} collisions at RHIC energies. The \mbox{Cu--Cu} data
at $\sqrt{s_{\rm NN}} = 62.4$~GeV shows
a small structure for non-central collisions whereas at higher energies the data are 
smoother. Dynamical fluctuations, expressed in terms of $N\nu_{+-dyn}$ are shown in
the right panel of Figure~\ref{mult_fluc_rhic} as function of collision centrality for
\mbox{Au--Au} collisions at $\sqrt{s_{\rm NN}} = 130$~GeV as measured by
the STAR experiment\cite{star_mult}.  The open circles show the
measured data compared to the charge conservation limit (dotted line), resonance gas (solid line) and HIJING calculations (solid squares).
Detailed understanding of these results would require considerations of
centrality selection and detector effects. 

\section{Temperature and $\ave{\pt}$ Fluctuations}

The $\ave\pt$ of emitted particles in an event is related
to the temperature of the system. Thus the event-by-event fluctuations of
average $\pt$~ is sensitive to the temperature fluctuations predicted for 
the QGP phase transition. Several measures of fluctuation
have been introduced in order to probe the dynamical fluctuation from the
measured values, some of these include\cite{ray02}:
\begin{eqnarray}
F_{\pt} & = & \frac{\Omega_{data} - \Omega_{baseline}}{\Omega_{baseline}}, \\
{\rm where ~~ } \Omega & = & \sigma_{M_{\pt}}/\ave{N}, \\ 
\Delta\sigma^2_{\pt} & \equiv & \frac{1}{\varepsilon} \sum_{j=1}^{\varepsilon}
N_j \left( \langle \pt \rangle_j - \overline{\pt} \right)^2
- \sigma_{\hat{p}_T}^2
\equiv 2 \sigma_{\hat{p}_{\rm T}} \Delta\sigma_{\pt}, \\
\Phi_{\pt} & \equiv & \left[ \frac{1}{\varepsilon} \sum_{j=1}^{\varepsilon}
\frac{N^2_j}{\langle N \rangle} ( \langle \pt \rangle_j - \overline{\pt} )^2 
 \right]^{1/2} - \sigma_{\hat{p}_{\rm T}}. \\
\sigma^2_{\langle \pt \rangle,{\rm dynamical}} & \equiv &
\frac{1}{\varepsilon} \sum_{j=1}^{\varepsilon}
\frac{1}{N_j (N_j - 1)}
\sum_{i \neq i^{\prime} = 1}^{N_j} \delta p_{{\rm T}_{ji}} \delta p_{{\rm T}_{ji^{\prime}}},
\end{eqnarray}
where $\varepsilon$ is the number of events, $j$ is the event index,
$N_j$ is the event multiplicity, $\langle N \rangle$ is the mean multiplicity,
$i$ is a particle index, and $\delta p_{T_{ji}} = p_{T_{ji}} - \overline{\pt}$.
For minimal variations of $N_j$ within the event ensemble, one can define: 
\begin{eqnarray}
\Delta\sigma_{\pt} & \cong & \Phi_{\pt} \cong \frac{\langle N \rangle - 1}
{2 \sigma_{\hat{p}_{\rm T}}} \sigma^2_{\langle \pt \rangle,{\rm dynamical}} \\
{\rm and ~~} \ave {\Delta p_{i,1} \Delta p_{i,2}} & = & \frac{1}{N_{\rm event}}
\sum_{k=1}^{N_{\rm event}} \sum_{j=1,i\ne j}^{N_{k}} 
\frac {\delta p_{{\rm T},j}\delta p_{{\rm T},i}}
{N_k (N_k-1)} 
\end{eqnarray}

 \begin{figure} 
  \begin{center}
    \leavevmode
    \epsfig{file=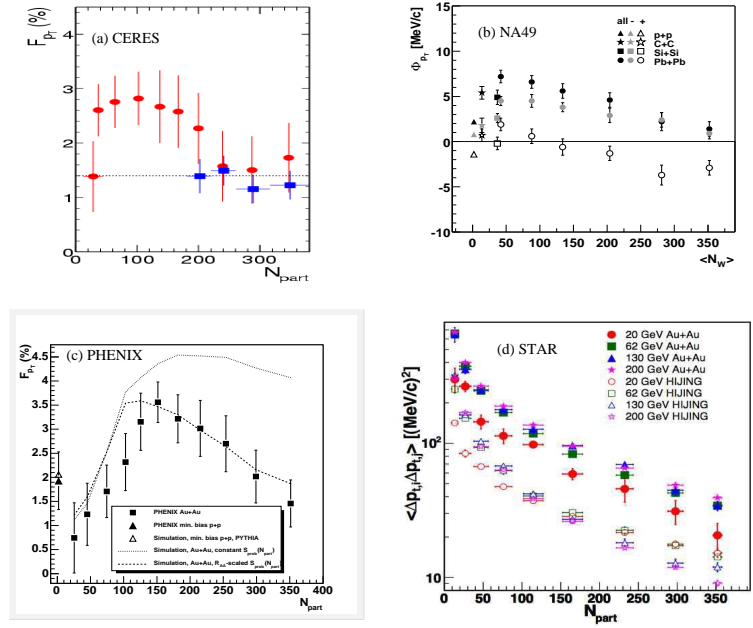,width=4in}
  \end{center}
\vspace*{-0.5cm}
  \caption{Dynamical $\ave\pt$ fluctuations as a function of
centrality of the collision from (a) CERES, (b) NA49, 
(c) PHENIX and (d) STAR experiments.
}
  \label{pt_fluc}
\end{figure}

Figure~\ref{pt_fluc} shows 
the centrality dependence of dynamical fluctuations reported by
CERES\cite{ceres_pt}, NA49\cite{na49_pt},   
PHENIX\cite{phenix_pt} and STAR\cite{star_pt} experiments. The results
presented in Figure~\ref{pt_fluc}(d) show a smooth variation of
fluctuation with centrality whereas the other measurements show
non-monotonic behaviour. Efforts are being made to understand
the nature and origin of these fluctuations. Because of the choice
of several variables, extraction of an excitation energy plot
combining data from SPS to RHIC is not straightforward. It is of
interest to us to have a common framework for presenting
the results from different experiments. 

\begin{figure}
  \begin{center}
    \leavevmode
    \epsfig{file=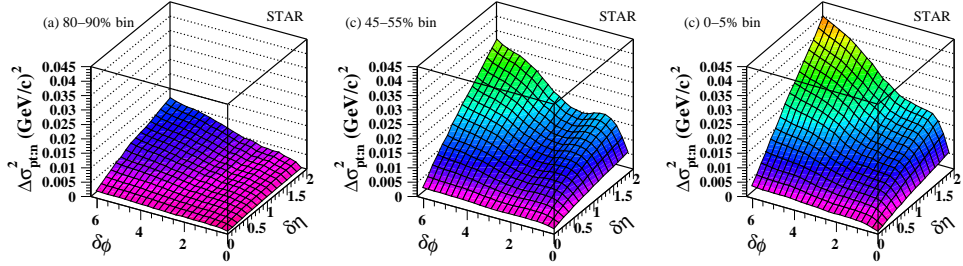,width=5in}
  \end{center}
\vspace*{-0.5cm}
  \caption{Scale dependence of $\ave{\pt}$ fluctuation within
the STAR acceptance expressed in terms of per-particle variance difference.}
  \label{pt_etaphi}
\end{figure}

In order to be
more sensitive to the origin of fluctuations, differential
measures have been adopted where the analysis is performed at
different scales (varying bins
in $\eta$ and $\phi$).
The scale dependence of $\ave{p_{\rm T}}$ fluctuation for
three centralities in \mbox{Au--Au} collisions at
$\sqrt{s_{\rm NN}}=200$~GeV \cite{star_pt_etaphi} is shown 
in Figure~\ref{pt_etaphi}. The extracted autocorrelations are
seen to vary rapidly with collision centrality, suggesting that
fragmentation is strongly modified by a dissipative medium in
more central collisions relative to peripheral collisions. Further
studies for different charge combinations will provide more
detailed information.

\section{Fluctuations in elliptic flow and eccentricity}

Fluctuations in physical quantities 
%that arise out of hydrodynamic predictions
can discern whether the matter created in heavy-ion collisions 
is a perfect fluid or not. A dissipation in
a non-perfect fluid is related to the fluctuations of the physical
quantities\cite{flow1,bhalerao}. Fluctuation in elliptic flow ($v_2$)
has been proposed to be a sensitive probe for this study, as it might
reflect the fluctuation in the initial spatial eccentricity.
Fluctuation in $v_2$ is argued to be also sensitive to
the following physical effects: (a) filamentation instability initiated due to the strong
momentum anisotropy of the partonic system, and the generation and subsequent explosions of the 
topological clusters and (b) multiplicity fluctuations. Thus,
study elliptic flow ($v_2$) on an event-by-event basis
is expected to provide sensitivity to initial conditions for the matter
created in heavy-ion collisions.

\begin{figure}
\begin{center}
\includegraphics[width=1.0\textwidth]{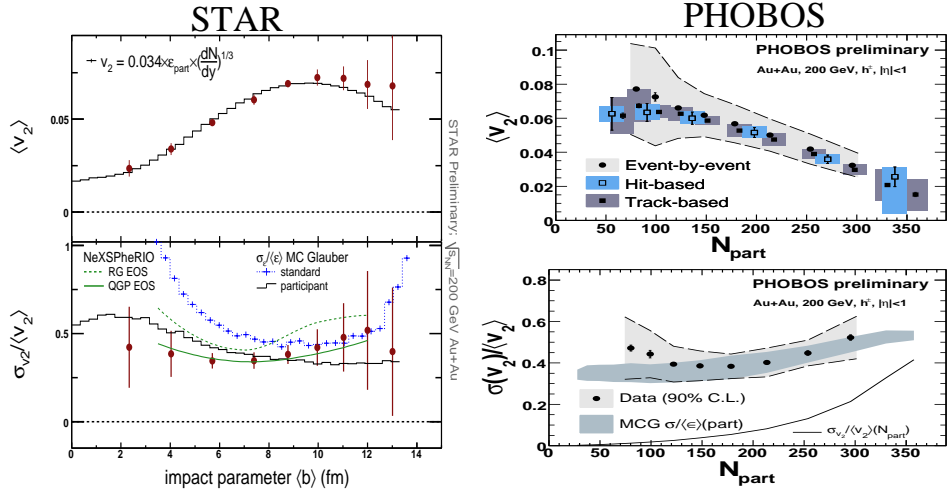}
\caption{
 The mean of the $v_2$ distribution ($\langle v_2\rangle$)
 (top panels) and the r.m.s. width of the the distribution
  ($\sigma_{v2}$) scaled by the mean (bottom panels) for \mbox{Au--Au} collisions 
  at $\sqrt{s_{\rm NN}} = 200$~GeV as measured by 
  for STAR and PHOBOS collaborations.
}
\label{flow}
\end{center}
\end{figure}

%\begin{figure}[htb]
%  \resizebox{0.47\textwidth}{0.305\textheight}{\includegraphics{star_data_vs_b.eps}}
%  \resizebox{0.50\textwidth}{!}{\includegraphics{data_vs_epart.eps}}
%  \resizebox{0.45\textwidth}{0.297\textheight}{\includegraphics{phobos_paperplot1.eps}}
%
%\caption[]{ 
%$\langle v_2\rangle$)
%  (top panels) and the r.m.s. width of the the distribution
%  ($\sigma_{v2}$) scaled by the mean (bottom panels). Data are
%  presented versus participant eccentricity (left panel) and impact
%  parameter (right panel). Various curves are explained in the
%  text. Impact parameter fluctuations have been removed so that the
%  data points represent the $v_2$ fluctuations for a fixed value of
%  $b$.}
%\label{f2}
%\end{figure}

Recently both STAR
and PHOBOS experiments have studied the fluctuations
in elliptic flow in \mbox{Au--Au} collisions at $\sqrt{s_{\rm NN}} = 200$~GeV.
The results are presented in Figure~\ref{flow}. The left panel shows the
STAR results\cite{star_flow_fluc} for mean ($\ave{v_2}$) and relative fluctuations ($\sigma_{v_2}/\ave{v_2}$)  
as a function of the impact parameter,
whereas the right panel shows the PHOBOS results\cite{phobos_flow_fluc} for the same quantities as
a function of the number of participants. The relative fluctuations have been
found to be about 36-40\%. The interesting fact is that these values can be
nicely reproduced by Monte-Carlo Glauber calculations of participant eccentricity,
implying that the later collision stages do not significantly alter the fluctuation pattern.
These results, along with results for other colliding systems and collision energies, will be able
to constrain the inputs to hydrodyanic model calculations.

\section{Event-by-event analysis of HBT radii}

\begin{figure}
\begin{center}
\includegraphics[width=0.4\textwidth,angle=270]{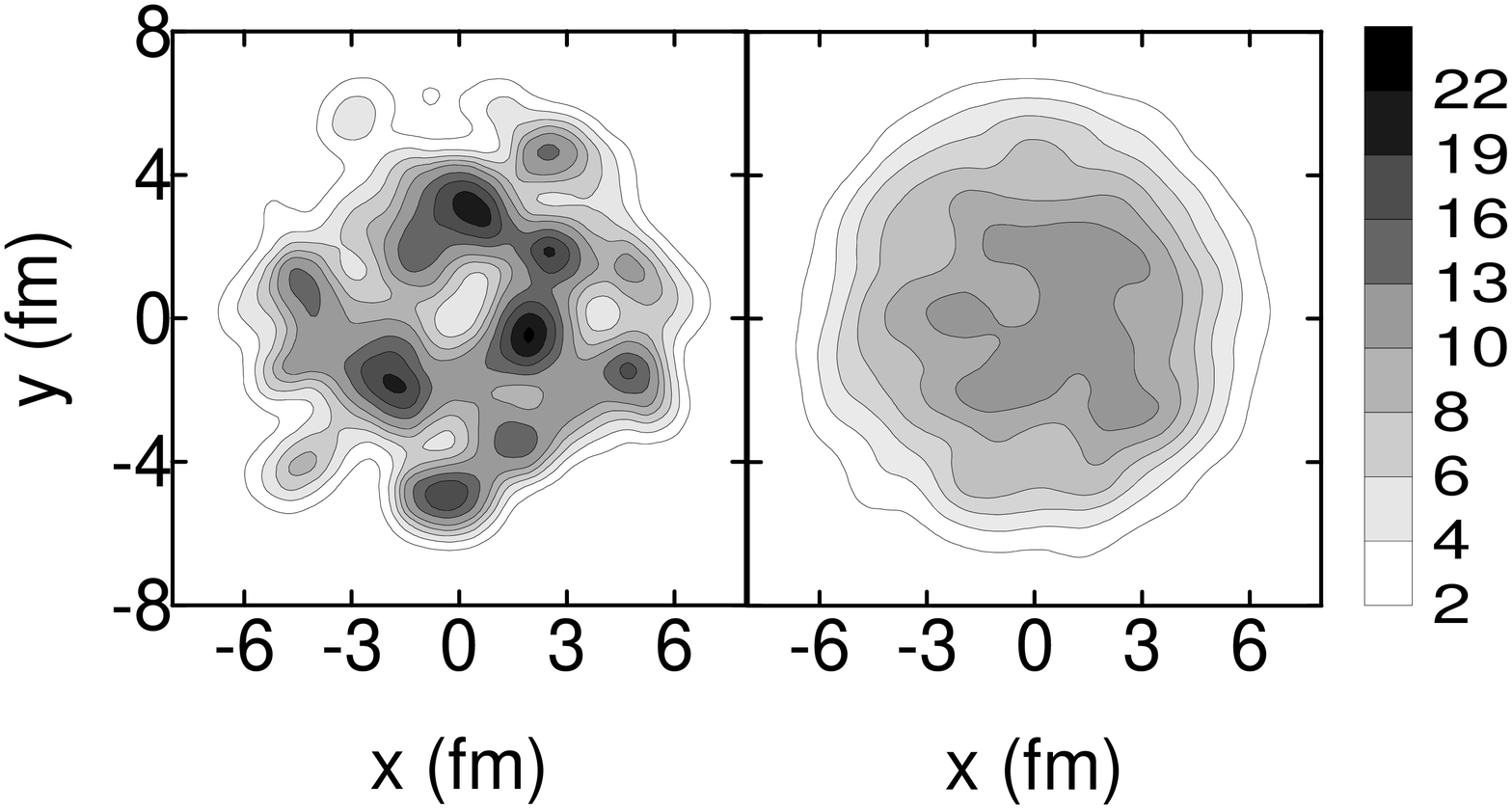}
\caption{
Energy density distributions (in units of GeV/fm$^3$) 
plotted in $x-y$ source dimensions for a single event (left panel) and
average over 30 events. 
}
\label{HBT1}
\end{center}
\end{figure}

The information about the space-time structure of the emitting source can be
extracted by the method of intensity interferometry techniques, known as
Hanbury-Brown Twiss (HBT) correlations. Due to lack of statistics,
the analysis of HBT correlations
are performed over a large number of events. 
But in reality, the space-time structure
of the emitting source may vary from one event to other.
This is illustrated in Figure~\ref{HBT1}, where the energy density distribution
is plotted in $x-y$ source dimensions\cite{fluc_hbt}. The distribution for
a typical single event (left panel)
shows several blobs of high density matter, 
whereas the distribution is smoothed out if an
average is taken (right panel).
It would be interesting to
perform correlation function analysis for single events from which one can
understand fluctuations in three dimensional source sizes. These fluctuations
will provide important information about the initial source sizes and could
be related to initial eccentricity as well.

\begin{figure}
\vspace*{-0.4cm}
\begin{center}
\includegraphics[width=0.7\textwidth]{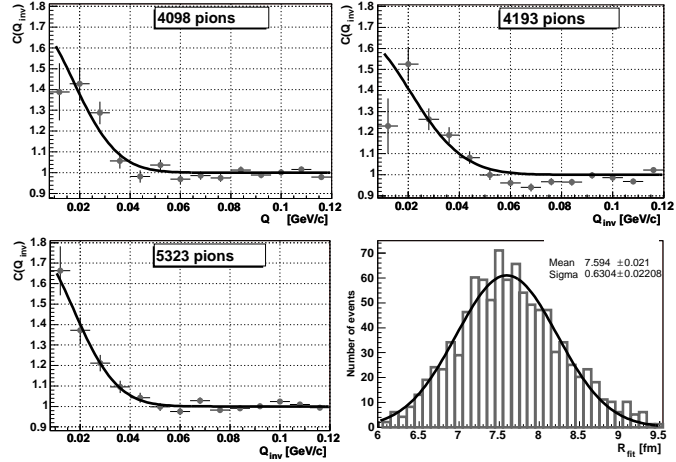}
\caption{
Single event correlation functions for three events with different number of reconstructed
pions. The bottom right panel shows the distribution of reconstructed radii 
%{\cite{alice_ppr}}.
}
\label{HBT2}
\end{center}
\end{figure}

An attempt has been made to perform single event HBT analysis 
for the simulated events corresponding \mbox{Pb--Pb} collisions at $\sqrt{s_{\rm NN}} = 5500$~GeV
in the framework of ALICE experiment{\cite{alice_ppr}} at the LHC. 
Figure~\ref{HBT2} shows the HBT correlations
for three typical events and a distribution of the reconstructed radii taken over
several events. The results indicate that it will be possible study
single-event interferometry in ALICE which
may for the first time be sensitive enough to source fluctuations.

%\section{Strangeness fluctuations}

\section{Fluctuation in particle ratio}

Relative production of different particle species produced in the hot and dense matter
might get affected when the system goes through a phase transition.
Of particular interest is the strangeness
fluctuation in terms of the ratio of kaons to pions. 
Large broadening in the yields of kaons to pions
has long been predicted because of the differences in free enthalpy of the
hadronic and QGP phase. This could be probed through the fluctuation in
the $K/\pi$ ratio.

\begin{figure}
\hspace{1pc}%
\begin{minipage}{14pc}
\includegraphics[width=14pc]{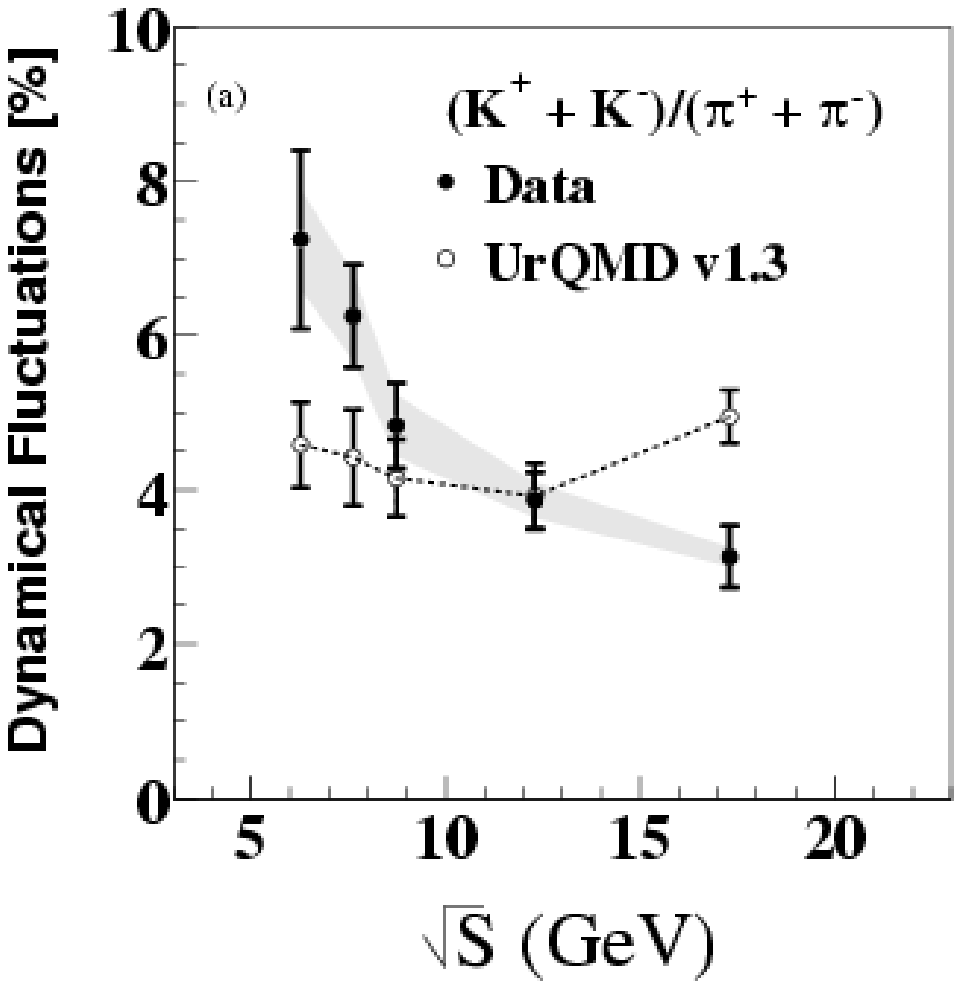}
\end{minipage}\hspace{0.0pc}%
\begin{minipage}{14pc}
\includegraphics[width=12pc]{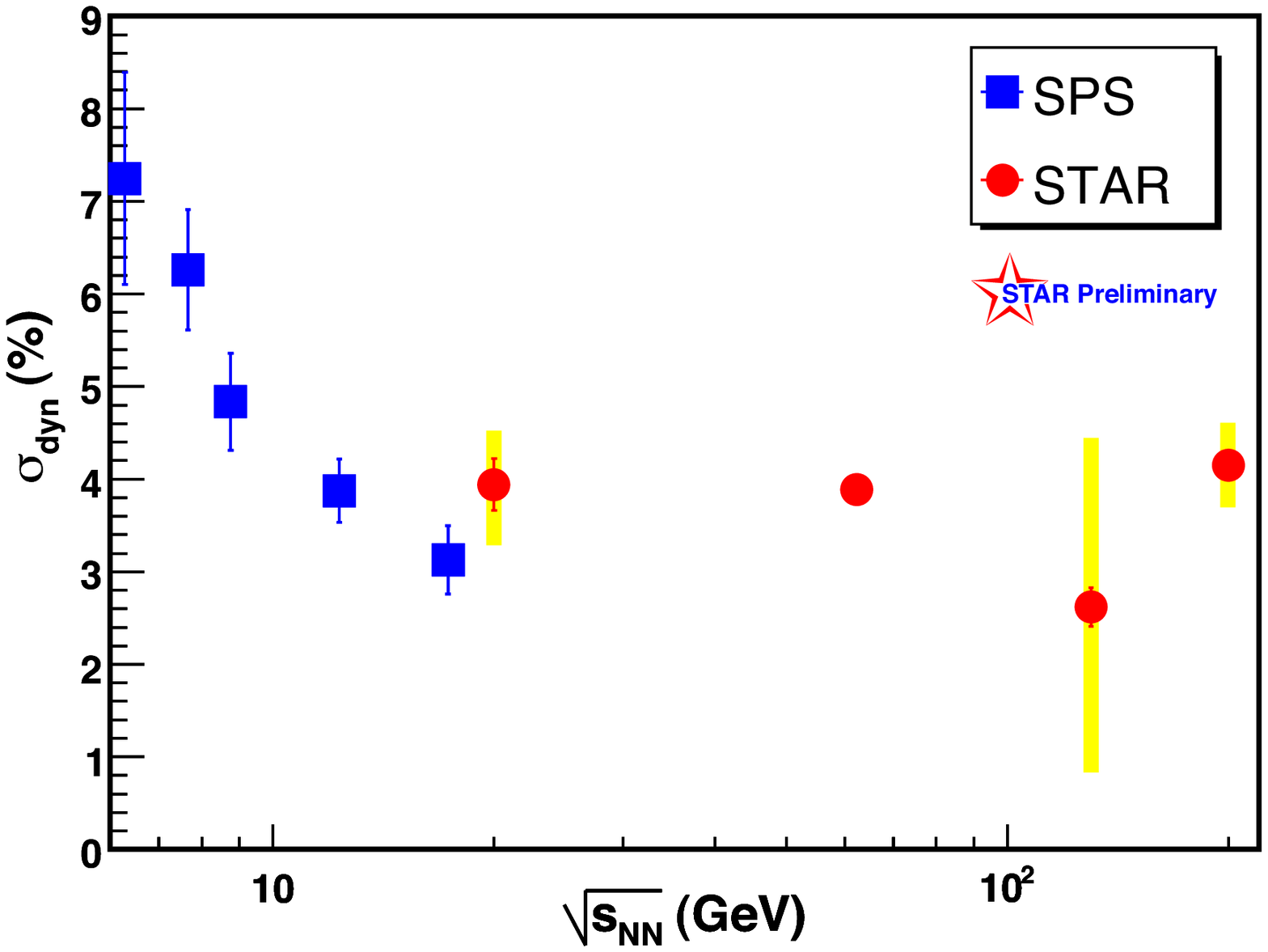}
\end{minipage} 
\caption{Excitation function for $\sigma_{dyn}$ of 
$[K^{+}+K^{-}]/[\pi^{+}+\pi^{-}]$ ratio at the SPS (left panel) and
with an extension to RHIC (right panel).}
\label{edep}
\end{figure}

A detailed study at SPS has been carried out at several beam energies\cite{qm04k2pi}. 
The ratio of inclusive mid-rapidity yields of
$\ave{K^-}/\ave{\pi^-}$ has an increasing trend with beam energy, whereas a
horn structure is seen in the ratio of  $\ave{K^+}/\ave{\pi^+}$. It has been shown that 
the dynamical fluctuations ($\sigma_{\rm dyn}$) in the ratio of $p/\pi$ has an increasing trend with 
respect to beam energy. This feature could be explained by model calculations.
At the same time $\sigma_{\rm dyn}$ in the $K/\pi$ ratio
is seen to decrease with beam energy, a behavior which could not be explained by the
same model. The $\sigma_{\rm dyn}$ values at SPS energies are shown in the left panel
of Figure~\ref{edep}. 
The STAR experiment has performed a similar study on the 
event-wise fluctuations of 
the $K/\pi$ ratio for \mbox{Au--Au} collisions at $\sqrt{s_{\rm NN}} = 62.4$~GeV and 
$\sqrt{s_{\rm NN}}=200$~GeV\cite{star_kpi}. 
A reduction as a function of
centrality is reported for the two energies. The right panel of Figure~\ref{edep}
shows an excitation energy plot for $K/\pi$ ratio
extended up to the highest RHIC energies.
The fluctuation decreases with increasing energy up to the highest SPS energy and 
remains constant at higher RHIC energies. Theoretical investigations\cite{torr,torr1} 
are underway to explain such behaviour.

\section{Net charge fluctuations}

\begin{figure} 
  \begin{center}
    \leavevmode
    \epsfig{file=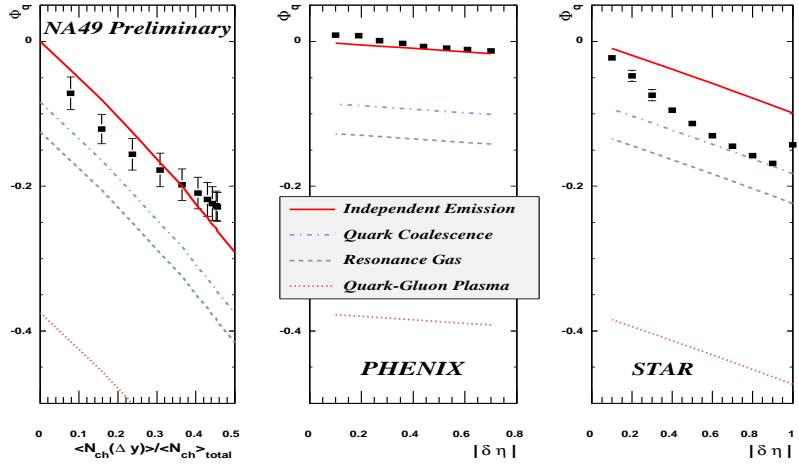, width=4.5in,height=3.1in}
  \end{center}
\vspace*{-1cm}
  \caption{Dynamical fluctuation of net charge for NA49, PHENIX and STAR 
  experiments.
}
  \label{netcharge_fluc}
\end{figure}

Fluctuations of conserved quantities like electric charge, baryon number or
strangeness are predicted to be significantly reduced in a QGP scenario
as they are generated in the early plasma stage of the system created in heavy-ion 
collisions with quark and gluon degrees of freedom\cite{As00,Ko00}.
The fluctuation generated at the QGP stage will increase 
as the system evolves in time\cite{evolution,Sh01}.
Net charge fluctuations have been measured by experiments at SPS and RHIC
using different fluctuation measures. 
Among these are $\Phi_q$ of NA49\cite{na49_net},
$\nu_{+-,dyn}$ of STAR\cite{star_mult} and $v(Q)$ as well as $\nu_{+-,dyn}$ 
used by PHENIX\cite{phenix_net}. A common framework 
which relates these variables has been used to compile
the available results\cite{volo,mitchell}. The results from these experiments are
shown in Figure~\ref{netcharge_fluc},
along with predictions from independent particle emission, quark coalescence, resonance
gas and a QGP scenario. Both NA49 and PHENIX results are consistent with the 
independent particle emission scenario, whereas the result for 
STAR is close to the case of the quark coalescence model. 

\section{Higher Moments of net charge}

\begin{figure}
\begin{center}
\includegraphics[width=5.9cm,height=4.7cm]{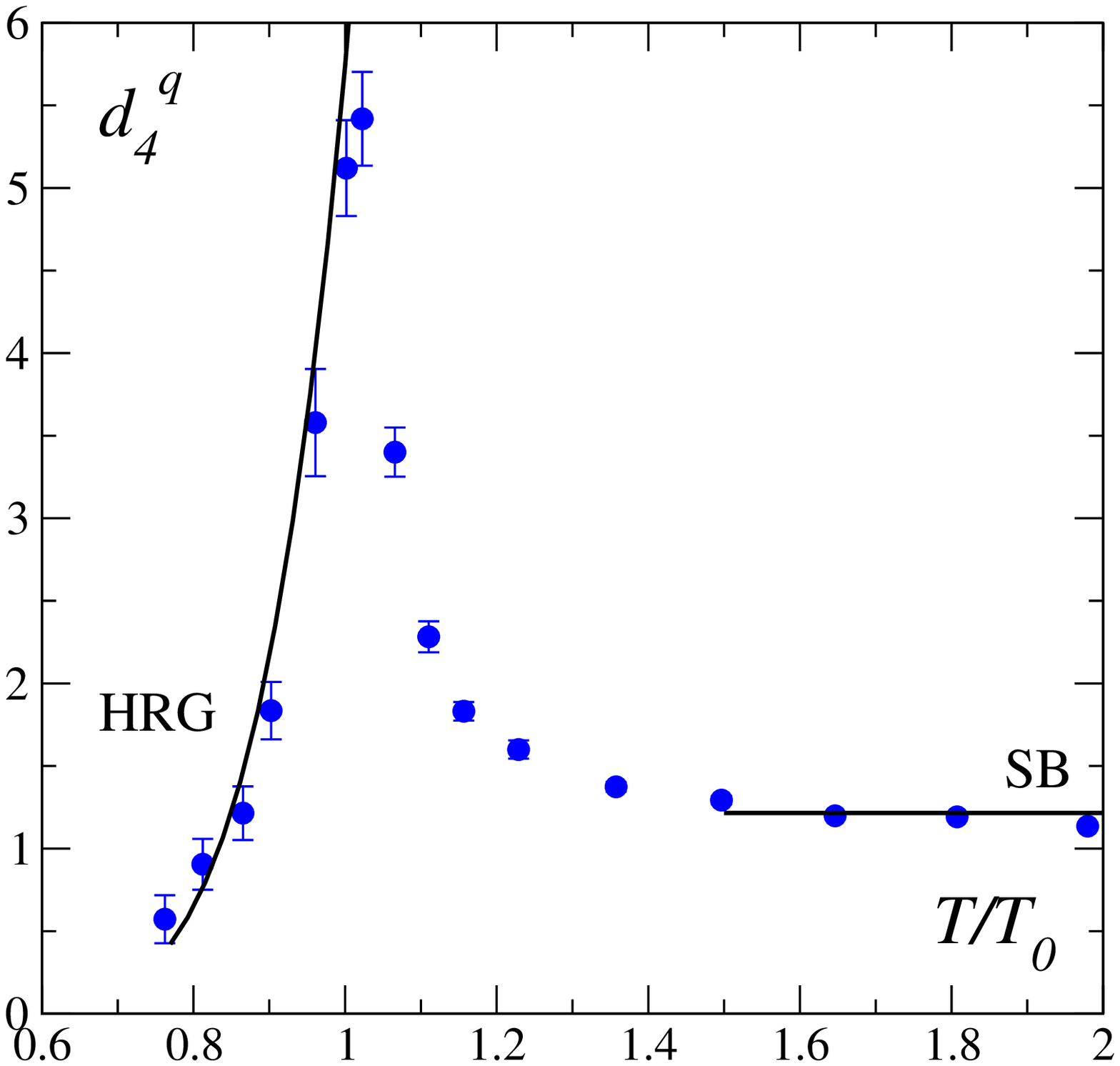}
\includegraphics[width=5.9cm,height=4.7cm]{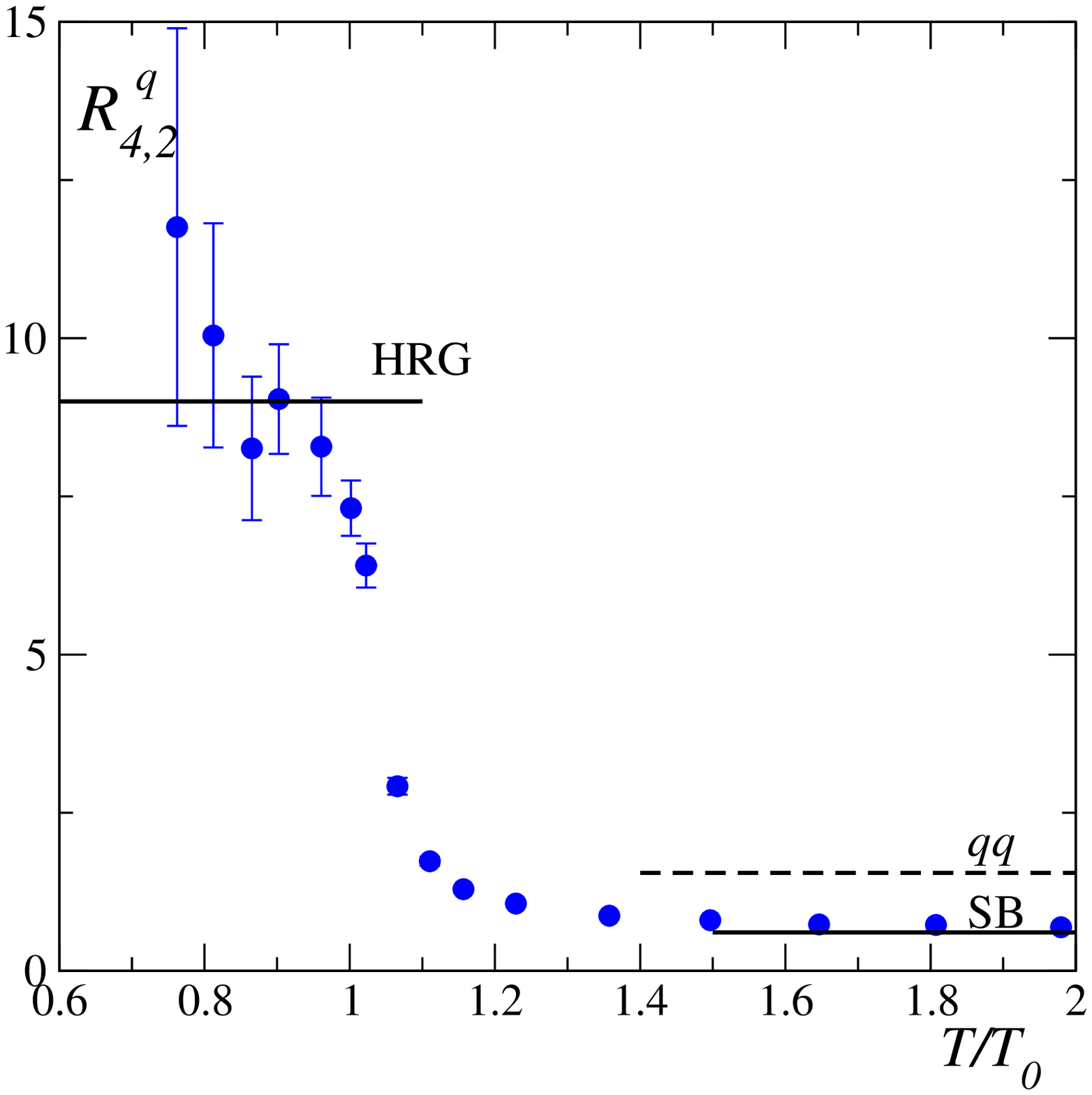}
\caption{The fourth-order cumulant moments of net charge from the lattice
calculations (left panel). The lines for $T/T_0<1$ are the hadron resonance gas model results.
The right panel shows the ratio of fourth to second order cumulants 
of quark number at $\mu_q=0$ 
}
\end{center}
\vspace*{0.5cm}
  \label{netcharge_moments1}
\end{figure}

Recently, lattice computations\cite{ekr,ggnls,redlich2006} have been performed
to study hadronic fluctuations. 
In the lattice framework one calculates the susceptibilities which are variances 
and covariances of various quantum numbers. These susceptibilities provide 
valuable information on the degrees of freedom in the hot phase of QCD.  
The non-linear susceptibilities (NLS) have been calculated which are 
higher derivatives of the pressure with respect to the chemical potential. 
These calculations predict an enhancement of
fluctuation in the hadronic phase and suppression of fluctuations in the high
temperature phase of the QGP. A prominent structure in the
higher order moments of net charge distributions 
have been observed for temperatures close to the transition
temperature. Figure~\ref{netcharge_moments1} shows the 4th order cumulants 
of the net charge and the 
ratio of the second to the fourth order cumulants of the net charge distributions\cite{redlich2006}. In the
hadronic phase this ratio has an increase with increasing temperature up to
the critical temperature, $T_C$, and a rapid suppression
is seen in the high temperature phase of QGP.

\begin{figure} 
\vspace*{-1.5cm}
  \begin{center}
    \leavevmode
    \epsfig{file=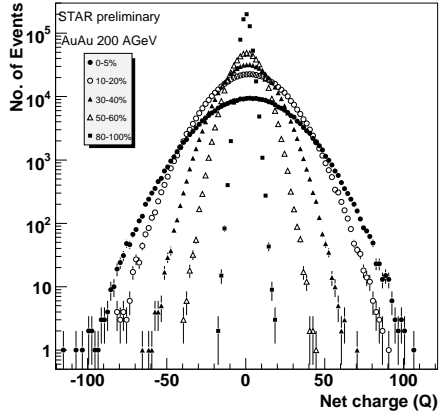, width=2.5in}
  \end{center}
\vspace*{-0.5cm}
  \caption{Net charge distributions of particles with $p_{\rm t}$ below 1GeV/c
for Au--Au collisions at $\sqrt{s_{\rm NN}} = 200$~GeV
at different centralities. 
}
  \label{netcharge_moments2}
\end{figure}

While the origin of this structure is under discussion by various authors, 
it provides an excellent opportunity for experiments to make a study. 
Figure~\ref{netcharge_moments2} shows the 
net charge distributions of particles
with $p_{\rm T}$ below 1GeV/c for Au--Au collisions at $\sqrt{s_{\rm NN}} = 200$~GeV for 
different centralities in the STAR experiment within a pseudorapidity coverage
of $-1\le \eta \le 1$. Efforts are underway to study higher order moments of
these distributions by making smaller bins in detector acceptances and $p_{\rm T}$. 
The ratio of the second to fourth order moments,
can be expressed in terms of the kurtosis of the net charge distributions.
This can provide a measure of the
deviation from a normal distribution in terms of
its peakiness (positive kurtosis) or flatness (negative kurtosis) at the mean.
Detailed studies are being performed.

\section{Balance functions}

The method of Balance Functions (BF)\cite{Pratt},
provides a measure of correlation of oppositely charged particles produced 
in heavy-ion collisions. The basic idea is that the charged hadrons are produced locally
as oppositely charged-particle pairs. The particles of such a pair are separated
in rapidity due to the initial momentum difference and secondary interaction with other
particles. The particles of a pair produced earlier are separated further in rapidity
compared to the particles coming from a pair produced later in time. Since the width of the 
correlation can be related to the time of hadronization of the charged particles,
this would signal any possible delayed hadronization, corresponding to QGP formation.

The BF can be studied as a function of several parameters in order to gain insight about 
different physics mechanisms. One of the basic studies may be performed in terms of
the relative pseudorapidity 
difference for all charged-particles.
In addition, there is the possibility to study the BF for different particle species
which could give insight to the different mechanisms that are important in the creation 
process for the species. Furthermore the BF can be studied as a 
function of the azimuthal angle\cite{Bozek}, $\phi$, and thus translate the 
correlation function into a measure of transverse flow. By doing that one will be able to 
quantify the transverse flow for different particle species.
The study of BF as a function of the invariant relative momentum $Q_{\rm inv}$
might yield a clear insight for interpreting the physics of the 
balancing charges.

\begin{figure}
\hspace{2pc}%
\begin{minipage}{12pc}
\includegraphics[width=12pc]{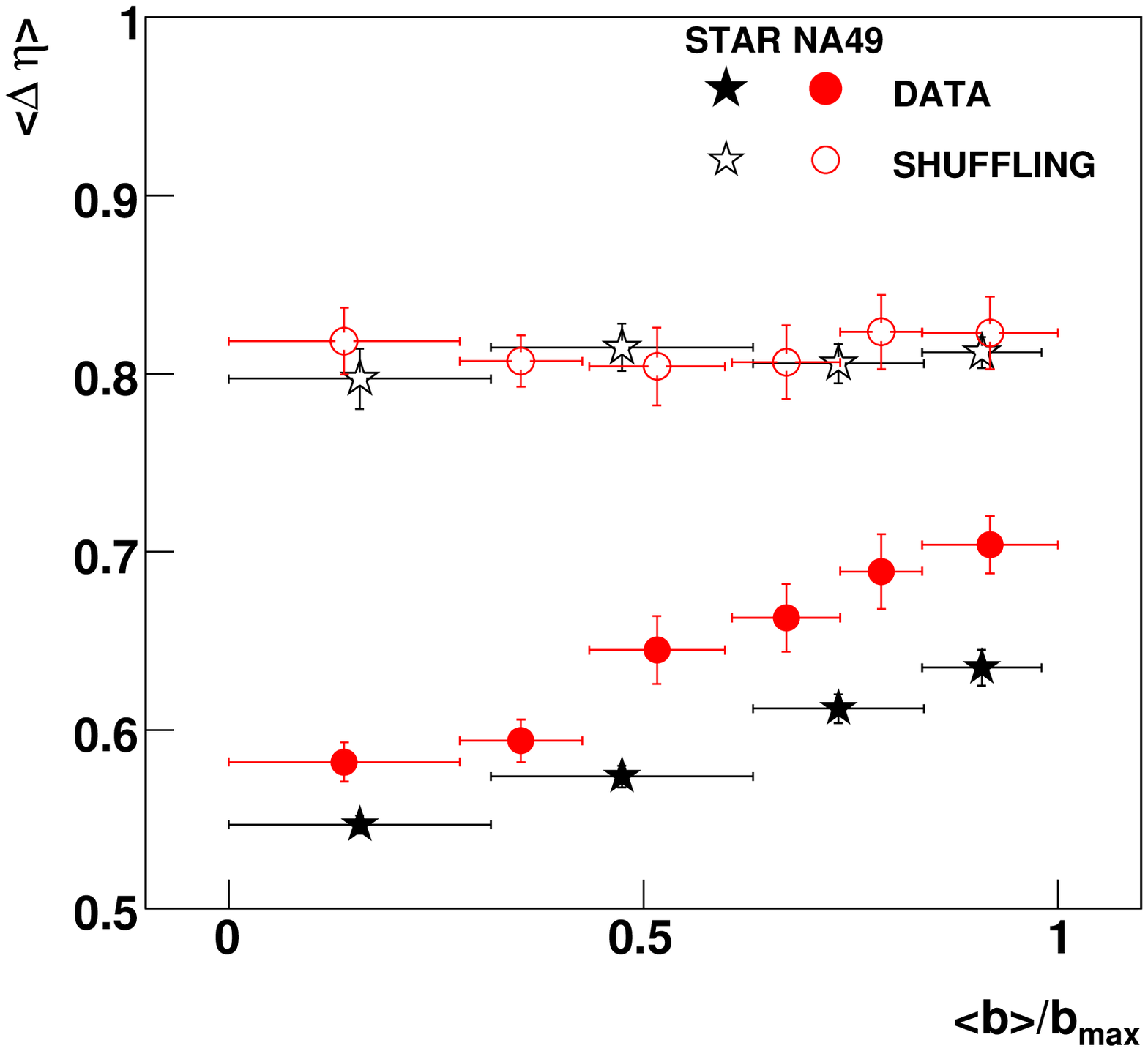}
\end{minipage}\hspace{0.1pc}%
\begin{minipage}{12pc}
\includegraphics[width=12pc]{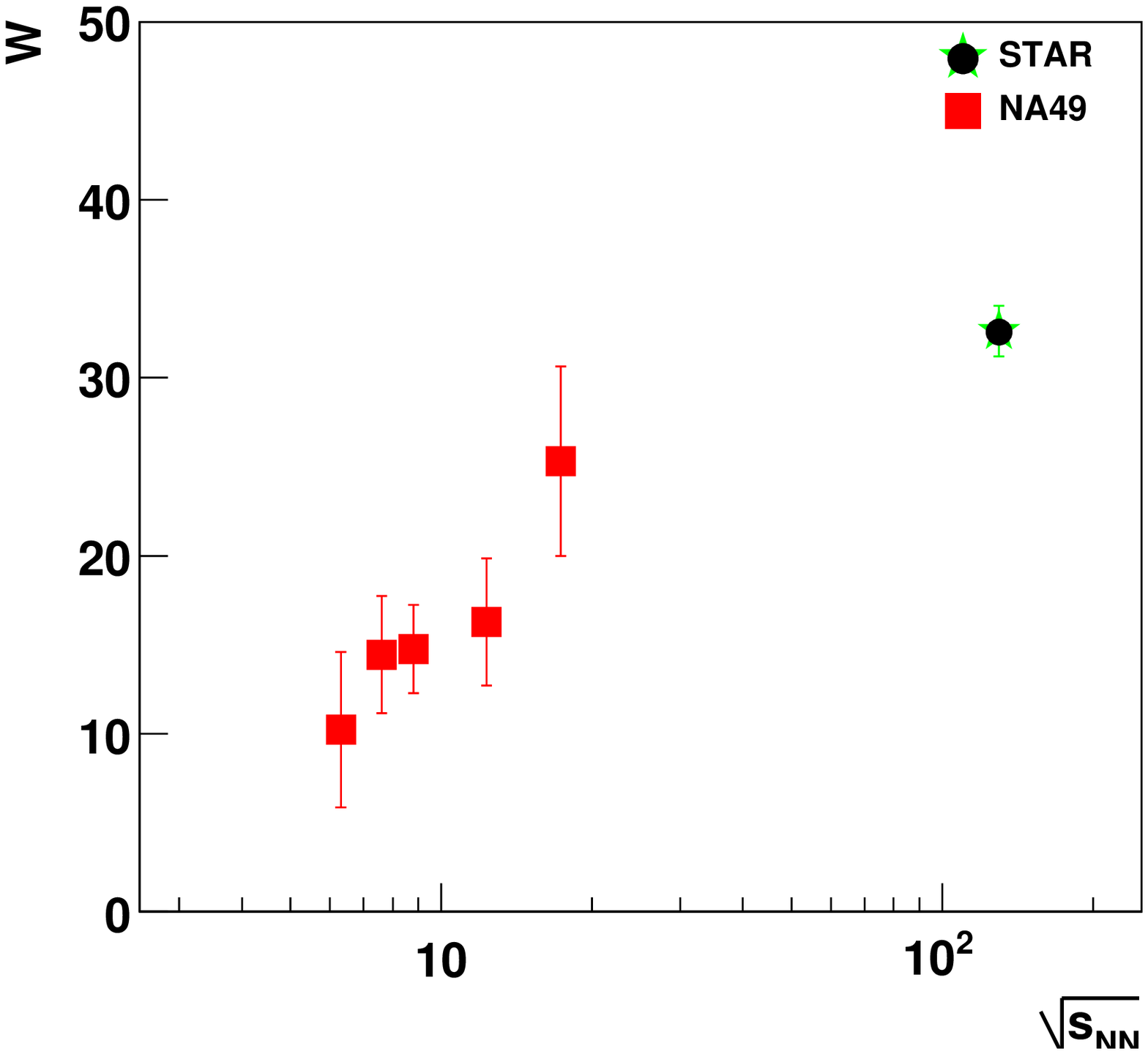}
\end{minipage} 
\caption{(a) The width of the balance function as a function of centrality for
experimental data along with results for shuffled bins (b) Normalized parameter ($W$) of
balance function as a function of beam energy.
}
\label{bf}
\end{figure}

Both STAR\cite{star_bf} and NA49\cite{na49_bf} experiments have
made detailed measurements of the BFs for
various colliding systems, centralities,
pseudorapidity intervals as well as for identified charged particles.
Here we present two of these studies; centrality dependence 
and excitation energy dependence of BF widths.
The left panel of Figure~\ref{bf} shows the width of the BFs as 
function of the normalized impact parameter for \mbox{Pb--Pb} collisions 
at $\sqrt{s_{\rm NN}}$=17.2~GeV and \mbox{Au--Au} collisions at 
$\sqrt{s_{\rm NN}}$=130~GeV.
The widths of the BF decrease from peripheral to central collisions 
in experimental data whereas the shuffled data shows no such reduction.
The decrease in the width can be quantified by the use of a normalized parameter, 
$W$, expressed as enhancement in the width in the data with respect to the
corresponding shuffled values.
The values of $W$ are plotted in the right panel of Figure~\ref{bf} as a function of
beam energy\cite{na49panos}. The increase of the $W$
from SPS to RHIC may be interpreted in terms of a delayed hadron scenario.

\section{Short and Long range correlations}

A copious production of partons, mainly gluons, due to hard 
and semi-hard processes, is expected in heavy-ion collisions.
During the early stages of collision the system is on average locally colourless,
but random fluctuations can break the neutrality\cite{capella}. 
Since the system 
is initially far from equilibrium, specific colour fluctuations can 
exponentially grow in time and then noticeably influence the
evolution of the system. 
Additional valuable information on the collision dynamics, specifically
on the string fusion and percolation phenomenon, may be
obtained in the event-by-event studies of
the correlations between various observables measured in separated rapidity
intervals (long range correlations).
These can be studied in different rapidity intervals for 
multiplicity correlations, $\ave{\pt}$ correlations and multiplicity-$\ave{\pt}$ correlations.
Model-independent detailed experimental information
on long-range correlations between such observables as charge, 
strangeness, multiplicity and $\ave{\pt}$ 
could be a powerful tool to discriminate theoretical reaction mechanisms.

\begin{figure}[htp]
\centerline{\includegraphics[width=6.1cm,height=4.7cm]{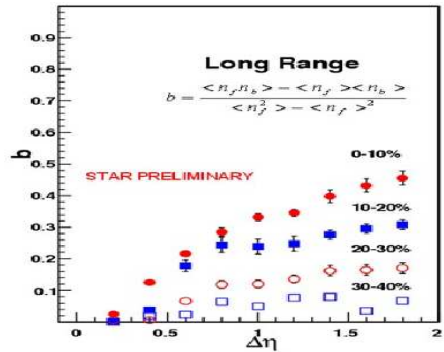}}
\caption{
Long range forward backward correlations
in rapidity measured in the STAR experiment. Short range correlations
have been subtracted.
}
\label{lrc}
\end{figure}
The results on the forward 
backward rapidity correlations measured as a function of centrality has been
reported by STAR\cite{terry,brijesh} and PHOBOS\cite{phobos} collaborations.
The results shown in Figure~\ref{lrc} from the STAR experiment shows the correlation 
to be quite strong, with an
increasing function of centrality of the collision. This could be qualitatively
understood in terms of long range longitudinal fields, such as in the Glasma or
string models\cite{mclleran}. An interesting analogy of the long range correlations could
be made with the 
amplification of quantum fluctuations to macroscopic magnitudes in the
early universe which form galaxies and clusters of galaxies\cite{mclleran}.

\section{Disoriented chiral condensates}

The QCD phase transition is predicted to be accompanied by chiral symmetry
restoration at high temperatures and densities. 
One of the most interesting consequences of chiral transition is the formation of a
chiral condensate in an extended domain, such that the direction of the condensate
is misaligned from that of the true vacuum. This phenomenon is termed as the
disoriented chiral condensates (DCC)\cite{Blai92,Bj93,raja93,dcc_phyrep}.
The formation of DCC results in an excess of low momentum pions in a single direction
in isospin space giving rise to large imbalances in the production of charged to neutral pions.
This is studied in terms of the distribution of neutral pion fraction, $f$, given by,
    \begin{equation}
       f=\frac{N_{\pi^0}}{N_\pi},
       \label{dcceqn}
    \end{equation}
where $N_{\pi^0}$ and $N_\pi$ are the number of neutral pions and total pions,
respectively. The pions in a normal event would follow a binomial form with a mean
of 1/3, whereas within a domain of DCC the probability of pion fraction would follow
a binomial distribution pattern such as, 
\[ P(f)=\frac{1}{2\sqrt{f}}. \]

The formation of DCC was hypothesized in the context of explaining observed 
abnormal events
from cosmic ray experiments\cite{La_80,Gl01} which had either excess
of charged-particles compared to neutrals (called Centauro events) or excess of neutrals
with respect to charged-particles (anti-Centauro events). 
A dedicated experiment, MiniMax, was set up at the Tevatron at Fermilab
to study p+$\bar{\rm p}$ collisions at $\sqrt{s}$ = 1.8~TeV\cite{Br:97}. 
At the SPS, both WA98 and NA49 experiments 
searched for the formation of DCC in heavy-ion 
collisions\cite{wa98_dcc1,wa98_dcc2,wa98_dcc3,na49_1,tapan}.

\begin{figure}[htp]
\centerline{\includegraphics[scale=0.45]{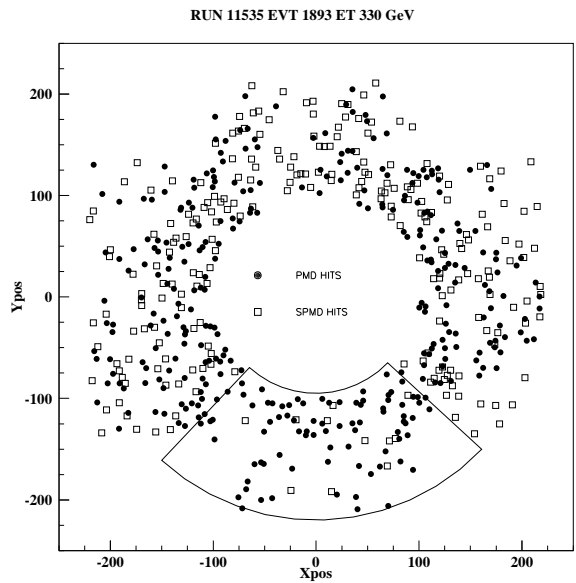}}
\caption{Photon (PMD) and charged-particle (SPMD)
hits in an azimuthal plane in the WA98 experimental set-up. 
The marked  $90^{\circ}$ patch corresponds to $f_{\rm max}$=0.77.
}
\label{mma}
\end{figure}

A thorough DCC search in PbPb~collisions at $\sqrt{s}_{\rm NN}$=17.2~GeV~
was performed by
the WA98 Collaboration at CERN.
This was based on a systematic study of photon and charged-particle multiplicity correlation
using the data from a preshower photon multiplicity detector (PMD) and a silicon pad
multiplicity detector (SPMD) for charged-particles. 
The analyses are performed using correlations of the number of photons to charged
particles, wavelet techniques and power spectrum analysis of
anomalous fluctuations in charged particles to photons. 
No clear DCC signal was observed and the upper limit
for DCC production at 90\% CL was established as a function of the fraction
of DCC pions among all pions produced.
An event display of the $x-y$ positions of charged particle (SPMD) and photon (PMD) hits 
is shown in Figure~\ref{mma} where a patch is marked which has a large number
of photons to charged particles. A sliding window analysis\cite{mma} method 
has been employed to identify such events for proper characterization.

\section{Fluctuations in the presence of jets}
The presence of jets and minijets may
affect the event-by-event fluctuation, which will be quite crucial
at LHC energies. In order to make any inference about
the fluctuation we need to understand the effect well.
On the other hand, this study may help in our understanding
of passage of jets through the medium.
A study of fluctuations in \pt~ has been made in the 
presence of jets for simulated events at LHC energies\cite{alice_ppr}.
The dependence of $\Phi_{\pt}$ on a search window defined by, 
$L_{\eta,\phi}=\sqrt{\Delta\eta^2+\Delta\phi^2}$ has been 
studied for soft particles and soft+hard particles.
The fluctuations seem to drastically increase in the
presence of hard particles when the window in terms of $L_{\eta,\phi}$ is increased. 
The expected jet production in \mbox{Pb--Pb} collisions at $\sqrt{s_{\rm NN}} = 5.5$~TeV for
LHC energies would lead to large EbyE fluctuations of $\ave{\pt}$. This may allow one to test
various models of jet production in the region not accessible by standard methods of jet
detection. On the other hand, fluctuations due to jet production should be taken
into account when considering the fluctuations due to other processes.
\begin{figure}[htp]
\centering
\includegraphics[width=0.5\textwidth]{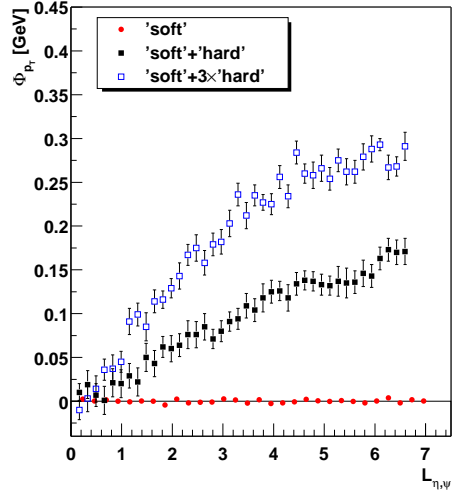}
\caption{
The dependence of $\Phi_{\pt}$ 
on the acceptance for
`soft' component (dots), `hard' + `soft' component (squares)
and the contribution of hard component
increased by a factor of 3 (open squares).}
\end{figure}

\section{Summary and outlook}

Experiments at SPS and RHIC have given a wealth of data on fluctuations of
various observables, some of the important ones have been discussed here.
The extraction of dynamical fluctuations originating from QGP
phase transition from the experimental results
becomes complicated because of several competing processes. 
We have attempted to understand the 
importance of proper centrality selection for fluctuation studies in
terms of a participant model. In order to infer any
dynamical fluctuation arising from various physics processes one has
to make sure that the fluctuations in number of participants are minimal.
Available results have been
discussed in terms of fluctuations in multiplicity, temperature and 
$\ave\pt$, elliptic flow, HBT radii, particle ratio, net charge
and higher moments of net charge distributions, balance functions, long range
correlations, formation of disoriented chiral condensates and presence of jets.
Differential measures are being adopted in order to gain insight to
the details of fluctuation. All these information have to be put together
in order to arrive at the final conclusion.

One of the most important aspects of QGP study is the location of the critical point.
It may be possible to access this experimentally by scanning the QCD phase diagram
in terms of baryon chemical potential and temperature. This can be accomplished 
by varying beam energies 
from about $\sqrt{s_{\rm NN}}$=5~GeV to 100~GeV. 
Such a program has recently been undertaken
at RHIC\cite{rhic}. Experiments at GSI\cite{CBM} are planned to study this as well.
At higher energies of LHC
(\mbox{Pb--Pb} beams at $\sqrt{s_{\rm NN}}$=5500~GeV), the ALICE experiment 
will be able to make
precise event-by-event measurements of various quantities and study their fluctuations 
\cite{alice_ppr}. With continued development in new analysis methods and
theoretical advances, and with dedicated experiments, one will certainly
learn a great deal more about QGP phase transition through 
fluctuation studies.

%\section*{Acknowledgements}


\begin{thebibliography}{0}

\bibitem{jeon-koch} S. Jeon and V. Koch, {\it Quark gluon plasma}, Edited by R.C. Hwa and X.N. Wang,  (2003) 430, {\it Preprint} hep-ph/0304012. 

\bibitem{heiselberg}
    H. Heiselberg, Phys. Rep. {\bf 351} (2001) 161.

\bibitem{nayak}
    T.K. Nayak, Journal of Physics {\bf G32} (2006) S187-S194.
    {\it Preprint} nucl-ex/060802. 

\bibitem{vanhove} L. Van Hove, Z. Phys. {\bf C27} (1985) 135.

\bibitem{tricrit}
  M. A. Stephanov, K. Rajagopal and E. Shuryak, Phys. Rev. Lett. {\bf 81} (1998) 4816.

\bibitem{heinz} U. Heinz, {\it Preprint} nucl-th/0512051. 

\bibitem{evolution}
    B. Mohanty, J. Alam, T.K. Nayak, Phys. Rev. {\bf C67} (2003) 024904.

\bibitem{wa98_ebye}
    M.M. Aggarwal {\it et al.}, (WA98 Collaboration), Phys. Rev. {\bf C65} (2002) 054912.



\bibitem{baym}  
     G. Baym and H. Heiselberg, Phys. Lett. {\bf B469} (1999) 7.

\bibitem{venus}
     K. Werner, Phys. Rep. {\bf 232} (1993) 87.

\bibitem{na49_mult} C. Alt {\it et al.}, (NA49 Collaboration), {\it Preprint} nucl-ex/0612010 

\bibitem{phenix_mult} S.S. Adler {\it et al.}, (PHENIX Collaboration), {\it Preprint} nucl-ex/0409015.
\bibitem{phenix_mult_fluc}
     J. Mitchell {\it et al.} (PHENIX Collaboration), {\it Preprint} nucl-ex/0510076.

\bibitem{star_mult} 
         J. Adams {\it et al.}  (STAR Collaboration), Phys. Rev. {\bf C68} (2003) 044905.

\bibitem{ray02} R.L. Ray (STAR Collaboration) {\it Preprint} nucl-ex/0211030.

\bibitem{ceres_pt}
     Hiroyuki Sato {\it et al.} (CERES Collaboration), J. Phys. {\bf G30} (2004) S1371.

\bibitem{na49_pt} T. Anticic {\it et al.} (NA49 Collaboration),
       Phys. Rev. {\bf C70} (2004) 034902



\bibitem{phenix_pt}
     S.S. Adler {\it et al.} (PHENIX Collaboration), Phys. Rev. Lett. {\bf 93} (2004) 092301. 

\bibitem{star_pt} J. Adams {\it et al.} (STAR Collaboration) Phys. Rev. {\bf C72} (2005) 044902.

\bibitem{star_pt_etaphi} J. Adams {\it et al.} (STAR Collaboration) J. Phys. {\bf G32} (2006) L37.

\bibitem{flow1}
    S. Mrowczynski and E. Shuryak, Acta Phys. Pol. {\bf B34} (2003) 4241.

\bibitem{bhalerao} R.S. Bhalerao and J-Y Ollitrault, Physics Letters {\bf B641} (2006) 260.

\bibitem{star_flow_fluc} P. Sorensen (STAR Collaboration), {\it Preprint} nucl-ex/0612021.
\bibitem{phobos_flow_fluc} C. Loizides (PHOBOS Collaboration), {\it Preprint} nucl-ex/0701049.

\bibitem{fluc_hbt} O. Socolowski Jr. {\it et al.} Phys. Rev. Lett. {\bf 93} (2004) 182301.
\bibitem{alice_ppr} Physics Performance Report, Volume II, ALICE Collaboration, 
Journal of Physics {\bf G32} (2006) 1295.

\bibitem{qm04k2pi} C. Roland, J. Phys. {\bf G30} (2004) S1381.

\bibitem{star_kpi} S. Das {\it et al.} (STAR Collaboration), {\it Preprint} nucl-ex/0503023.

\bibitem{torr} G. Torrieri, S. Jeon and J. Rafelski, {\it Preprint} nucl-th/0510024

\bibitem{torr1} Georgio Torrieri,  Eur. Phys. Journal {\bf C49} (2007) 287.
     {\it Preprint} nucl-th/0702020 

\bibitem{As00} 
    M. Asakawa, U. Heinz, B.Muller, Phys. Rev. Lett. {\bf 85} (2000) 2072.

\bibitem{Ko00} 
    S. Jeon and V. Koch, Phys. Rev. Lett. {\bf 85} (2000) 2076.

\bibitem{Sh01}
    E. Shuryak and M.A. Stephanov, Phys. Rev. {\bf C63} (2001) 064903. 

\bibitem{na49_net} 
      C.~Alt {\it et al.}  (NA49 Collaboration), Phys. Rev. C\ {\bf 70} (2004) 064903.

\bibitem{phenix_net} 
      K.~Adcox {\it et al.}  (PHENIX Collaboration), Phys. Rev. Lett.  {\bf 89} (2002) 082301.

\bibitem{volo}
     C. Pruneau, S. Gavin, S. Voloshin, Phys. Rev. {\bf C66} (2002) 044904.

\bibitem{mitchell}
     J.T. Mitchell, J. Phys {\bf G30} (2004) S819.

\bibitem{ekr}
   S.\ Ejiri, F.\ Karsch and K.\ Redlich, Phys. Lett. B633 (2006) 275.
\bibitem{ggnls}
   R.\ V.\ Gavai and S.\ Gupta, Phys. Rev.  {\bf D72} (2005) 054006.
\bibitem{redlich2006} K. Redlich, B. Friman and C. Sasaki, {\it Preprint} nucl-ex/0702296.

\bibitem{Pratt} 
      S. A. Bass, P. Danielewicz and S. Pratt, Phys. Rev. Lett. {\bf 85} (2000) 2689.
\bibitem{Bozek} 
               P. Bozek, Phys. Lett. {\bf B609} (2005) 247.

\bibitem{star_bf} 
         J. Adams {\it et al.}, (STAR Collaboration) Phys. Rev. Lett. {\bf 90} (2003) 172301.
\bibitem{na49_bf} 
         C. Alt {\it et al.} (NA49 Collaboration), Phys. Rev. {\bf C71} (2005) 034903;
\bibitem{na49panos}
         P. Christakoglou {\it et al.} (NA49 Collaboration), {\it Preprint} nucl-ex/0510045.

\bibitem{capella} A. Capella {\it et al.} Phys. Rep. {\bf 236} (1994) 225.
\bibitem{terry}
        Terence Tarnowsky {\it et al.} (STAR Collaboration), {\it Preprint} nucl-ex/0606018
\bibitem{brijesh}
        Brijesh Srivastava {\it et al.} (STAR Collaboration), {\it Preprint} nucl-ex/0702042.
\bibitem{phobos}
         B.B. Back {\it et al.} (PHOBOS Collaboration) Phys. Rev. {\bf C74} (2006) 011901.
\bibitem{mclleran} Larry McLerran {\it Preprint} nucl-ex/0702004.

\bibitem{Blai92}
     J.-P. Blaizot and A. Krzywcki, Phys. Rev. {\bf D46} (1992) 246.
\bibitem{Bj93}
     J.D. Bjorken, K.L. Kowalski and C.C. Taylor, SLAC-PUB-6109, April 1993.
\bibitem{raja93}
     K. Rajagopal, F. Wilczek, Nucl. Phys. {\bf B399} (1993) 399.
\bibitem{dcc_phyrep}
      Bedanga Mohanty, Julien Serreau, Phys. Rep. {\bf 414} (2005) 263.

\bibitem{wa98_dcc1}
     M. M. Aggerwal {\it et al.} (WA98 Collaboration), Phys. Lett. {\bf B420} (1998) 169.

\bibitem{wa98_dcc2}
     M. M. Aggerwal {\it et al.} (WA98 Collaboration), Phys. Rev. {\bf C64} (2001) 011901.
\bibitem{wa98_dcc3}
     M. M. Aggerwal {\it et al.} (WA98 Collaboration), Phys. Rev. {\bf C67} (2003) 044901.
\bibitem{na49_1}
    H. Appelshauser {\it et al.} (NA49 Collaboration), Phys. Lett. {\bf B459} (1999) 679.

\bibitem{tapan}
     T.K. Nayak {\it et al.} (WA98 Collaboration), Nucl. Phys. {\bf A638} (1998) 249c.

\bibitem{La_80}
        C.M.G. Lates, Y. Fujimoto and S. Hasegawa, Phys. Rep. {\bf 65} (1980) 151
\bibitem{Gl01}
       E.G\l{}adysz-Dziadu\'s, INP Report No.1879/PH, Cracow, 2001.

\bibitem{Br:97}
       T.C. Brooks {\it et al.}, Phys. Rev. {\bf D55} (1997) 5667.

\bibitem{mma}
      M.M.Aggarwal {\it et al.} (WA98 Collaboration), Pramana {\bf 60} (2003) 987.

\bibitem{rhic}
        T. Ludlum {\it et al.} BNL-75692-2006, 
        Proceedings of the workshop on 
        ``Can we discover the QCD Critical point at RHIC?'', March 9-10, 2006.
\bibitem{CBM} 
        C. Hohne  {\it et al.} (CBM Collaboration) Nucl. Phys. {\bf A749} (2005) 141.

\end{thebibliography}
\end{document}